\documentclass[reprint,superscriptaddress,floatfix,amsmath,amssymb,aps,prl]{revtex4-1}

\usepackage{graphicx}
\usepackage{textcomp}
\usepackage{xcolor}
\usepackage{comment}
\usepackage[mathlines]{lineno}
\definecolor{light-gray}{gray}{0.8}
\definecolor{prlblue}{rgb}{0.176, 0.152, 0.57}

\usepackage[colorlinks=true, linkcolor=black, citecolor=prlblue, filecolor=black, urlcolor=prlblue]{hyperref}
\usepackage{textgreek}
\usepackage{lineno}  

\begin{document}


\title{Energy Depletion and Re-Acceleration of Driver Electrons \\ in a Plasma-Wakefield Accelerator}

\author{F.~Pe{\~n}a}
\email{felipe.pena@desy.de}
\affiliation{Deutsches Elektronen-Synchrotron DESY, Notkestr.~85, 22607 Hamburg, Germany}
\affiliation{Universit{\"a}t Hamburg, Luruper Chaussee 149, 22761 Hamburg, Germany}
\author{C.~A.~Lindstr{\o}m}
\affiliation{Deutsches Elektronen-Synchrotron DESY, Notkestr.~85, 22607 Hamburg, Germany}
\affiliation{Department of Physics, University of Oslo, 0316 Oslo, Norway}
\author{J.~Beinortait{\.e}}
\affiliation{Deutsches Elektronen-Synchrotron DESY, Notkestr.~85, 22607 Hamburg, Germany}
\affiliation{University College London, Gower St., London WC1E 6BT, United Kingdom}
\author{J.~Bj{\"o}rklund Svensson}
\affiliation{Deutsches Elektronen-Synchrotron DESY, Notkestr.~85, 22607 Hamburg, Germany}
\author{L.~Boulton}
\affiliation{Deutsches Elektronen-Synchrotron DESY, Notkestr.~85, 22607 Hamburg, Germany}
\affiliation{SUPA, Department of Physics, University of Strathclyde, Glasgow G4 0NG, United Kingdom}
\affiliation{The Cockcroft Institute, Daresbury, United Kingdom}
\author{S.~Diederichs}
\affiliation{Deutsches Elektronen-Synchrotron DESY, Notkestr.~85, 22607 Hamburg, Germany}
\affiliation{Universit{\"a}t Hamburg, Luruper Chaussee 149, 22761 Hamburg, Germany}
\author{B.~Foster}
\affiliation{Deutsches Elektronen-Synchrotron DESY, Notkestr.~85, 22607 Hamburg, Germany}
\affiliation{John Adams Institute for Accelerator Science at University of Oxford, Oxford, UK}
\author{J.~M.~Garland}
\affiliation{Deutsches Elektronen-Synchrotron DESY, Notkestr.~85, 22607 Hamburg, Germany}
\author{P.~Gonz{\'a}lez Caminal}
\affiliation{Deutsches Elektronen-Synchrotron DESY, Notkestr.~85, 22607 Hamburg, Germany}
\affiliation{Universit{\"a}t Hamburg, Luruper Chaussee 149, 22761 Hamburg, Germany}
\author{G.~Loisch}
\affiliation{Deutsches Elektronen-Synchrotron DESY, Notkestr.~85, 22607 Hamburg, Germany}
\author{S.~Schr{\"o}der}
\author{M.~Th{\'e}venet}
\author{S.~Wesch}
\author{J.~C.~Wood}
\author{J.~Osterhoff}
\author{R.~D'Arcy}
\affiliation{Deutsches Elektronen-Synchrotron DESY, Notkestr.~85, 22607 Hamburg, Germany}



\begin{abstract}
    For plasma-wakefield accelerators to fulfil their potential for cost effectiveness, it is essential that their energy-transfer efficiency be maximized.
    A key aspect of this efficiency is the near-complete transfer of energy, or depletion, from the driver electrons to the plasma wake.
    Achieving full depletion is limited by the process of re-acceleration, which occurs when the driver electrons decelerate to non-relativistic energies, slipping backwards into the accelerating phase of the wakefield and being subsequently re-accelerated. 
    Such re-acceleration is unambiguously observed here for the first time.
    At this re-acceleration limit, we measure a beam driver depositing (57~$\pm$~3)\% of its energy into a 195-mm-long plasma.
    Combining this driver-to-plasma efficiency with previously measured plasma-to-beam and expected wall-plug-to-driver efficiencies, our result suggests that plasma-wakefield accelerators can in principle reach or even exceed the energy-transfer efficiency of conventional accelerators.
\end{abstract}

\maketitle

The cost of particle-accelerator facilities for high-energy physics and photon science is increasing significantly~\cite{Livingston_1954, Shiltsev_PhysToday_2020}, driven by the need for higher particle energies and a limited accelerating gradient; conventional radio-frequency accelerators are limited to around 100~MV/m~\cite{Grudiev_PRSTAB_2009}.
Plasma-wakefield accelerators can exceed this by several orders of magnitude, providing accelerating gradients beyond 10~GV/m.
Such gradients can be sustained in the charge-density wake driven by an intense laser pulse~\cite{Tajima_PRL_1979,Esarey_RMP_2009} or charged-particle beam~\cite{Chen_PRL_1985,Ruth_PA_1985,Hogan_RAST_2017} (known as a \textit{driver}) propagating in a plasma.
In a plasma-wakefield accelerator, plasma electrons are expelled from the driver-propagation axis and subsequently attracted by the exposed slow-moving plasma ions, resulting in a region co-propagating with the driver that is both strongly focusing and accelerating for a trailing electron bunch~\cite{Rosenzweig_PRA_1991}.
While plasma wakefields promise to reduce dramatically the size and construction cost of future accelerators, high-power applications such as hard-x-ray free-electron lasers~\cite{Madey_JAP_1071,McNeil_NatPhoton_2010} and linear colliders~\cite{ILC_2013,CLIC_2012} also require highly energy-efficient acceleration.
This high efficiency ensures that running costs are minimized: a key requirement for sustainable and affordable operation and reduced environmental impact.

In a plasma-wakefield accelerator, energy transfer from the wall plug to the accelerated beam can be divided into three steps: firstly, wall-plug power is used to produce a driver; secondly, the driver energy is deposited into the plasma by creating a wakefield; lastly, the wake energy is extracted by accelerating a trailing bunch.
The efficiency of the first step is currently determined by the type of driver. 
For laser drivers, the technology to drive strong wakefields (Ti:sapphire lasers~\cite{Moulton_JOptSocAmB_1986}) has sub-percent-level wall-plug-to-driver efficiency~\cite{Hooker_JPhysB_2014}. Research on thin-disk~\cite{Nubbemeyer_OptLett_2017}, fiber~\cite{Jauregui_NatPhoton_2013} and Thulium~\cite{Tamer_OptExpress_2022} lasers may significantly improve this in the future. 
For particle-beam drivers---and in particular \textit{electron-beam drivers}---the production sources benefit from decades of development in klystrons~\cite{Baikov_IEEETransElectronDevices_2015} and conventional accelerating structures.
As an example, the CLIC linear-collider proposal has a wall-plug-to-driver efficiency as high as 55\%~\cite{CLIC_2012}. 
For the remainder of this Letter, we report on results for particle-beam drivers. 
In the second step, where the plasma wakefield is driven, the electrons of the driving beam are relativistic and thus their velocity changes negligibly until they are decelerated to non-relativistic energies (or \textit{energy depleted}).
Thus, the wakefield of a driver in a preformed plasma (i.e., negligible head erosion~\cite{Blumenfeld_Thesis_2009}), as experienced by a trailing electron bunch, remains approximately unchanged until depletion occurs. 
Finally, in the third step, to extract the wakefield energy with high efficiency, the wakefield created by the trailing bunch must be strong enough to destructively interfere with that of the driver with similar amplitude---a process known as \textit{beam loading}~\cite{vanDerMeer_CLICNote_1985,Katsouleas_PA_1987}.
Experiments have reached plasma-to-beam efficiencies as high as ($42\pm4$)\%~\cite{Litos_Nature_2014, Lindstroem_PRL_2021}.
While large energy loss for a few particles has been shown~\cite{Blumenfeld_AIPCP_2009,Blumenfeld_Thesis_2009}, the limit of driver energy depletion has not been experimentally investigated, nor has the driver-to-plasma efficiency been accurately quantified when operating at this limit.

\begin{figure*}[t]
	\centering\includegraphics[width=1\textwidth]{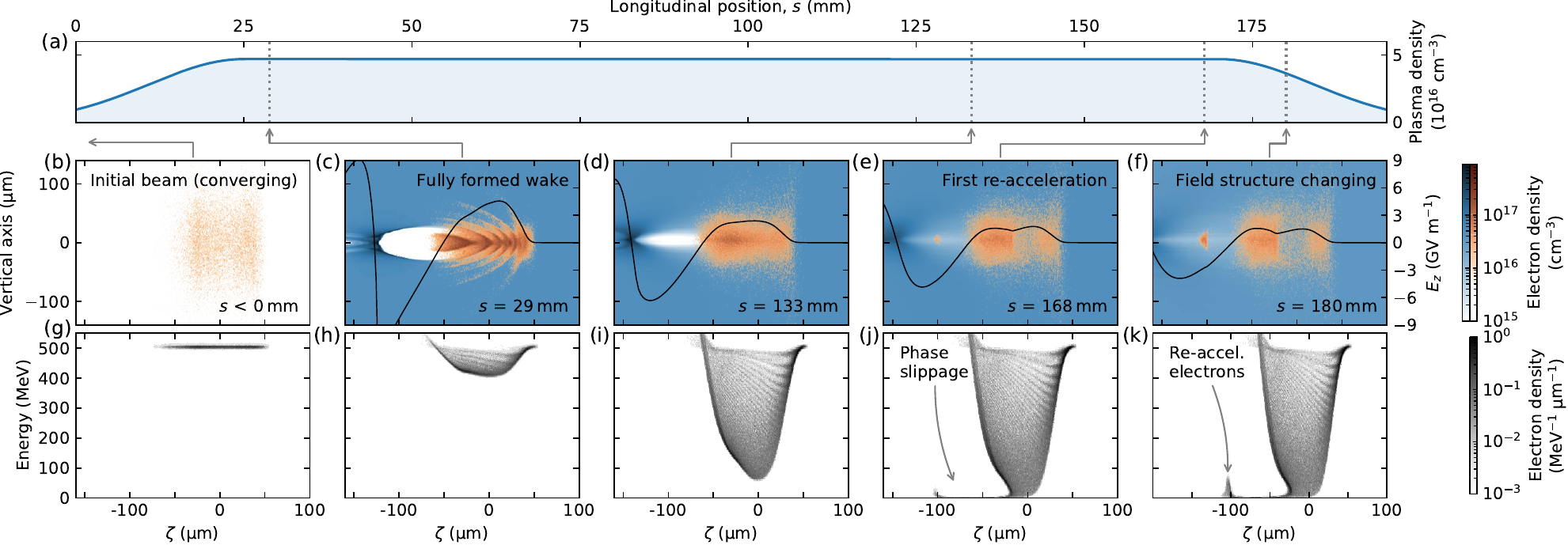}
	\caption{Snapshots from a PIC simulation showing driver energy depletion and electron re-acceleration. (a) The longitudinal plasma-density profile of the experiment is approximated with a central flattop, inner Gaussian ramps and an external exit ramp, as in Ref.~\cite{Lindstrom_NatCommun_2024}. (b--f) Density of plasma and beam electrons (blue and orange color maps, respectively) and the on-axis longitudinal electric field $E_z$ (black line). The corresponding longitudinal phase spaces are shown in (g--k). Driver electrons lose energy while locked in phase (g--i) until the electrons are depleted in energy and become non-relativistic, at which point they slip backwards in phase (j) and are re-accelerated (k). 
    $\zeta$ denotes the position along the bunch in a co-moving frame.}
    \label{fig:1_simulation}
\end{figure*}

Energy depletion of a beam driver is limited by electron \textit{re-acceleration} \cite{Jakobsson_PPCF_2019}.
In this process, for which there may be possible indications in Ref.~\cite{Chou_PhysRevLett_2016}, driver electrons decelerated to non-relativistic energies slip backward into the accelerating phase of the plasma wakefield, where they are accelerated back to relativistic energies and again become phase-locked with the driver.
Figure~\ref{fig:1_simulation} shows a particle-in-cell (PIC) simulation representative of the experimental setup, performed with the quasi-static code HiPACE++ \cite{Diederichs_2022}, which demonstrates this process.
The quasi-static approximation significantly lowers the computational cost \cite{Mora_PhysPlasmas_1997}, while accurately capturing re-acceleration (see the Supplementary Material).
Once the re-acceleration process starts, a steady stream of electrons from the decelerating to the accelerating phase reduces the charge driving the wakefield and consequently alters its structure [Fig.~\ref{fig:1_simulation}(e)].
Simultaneously, the re-accelerated charge causes beam loading, which changes the shape of the accelerating field [Fig.~\ref{fig:1_simulation}(f)].
If the longitudinal field experienced by the trailing bunch, averaged over the acceleration length, is not kept uniform (by way of precisely controlled beam loading \cite{Tzoufras_PRL_2008,Lindstroem_PRL_2021}), it will eventually increase the energy spread of the trailing bunch. 
Moreover, the presence of non-relativistic driver and potentially plasma electrons (from driving a weaker wake) within the trailing bunch can alter the focusing field through space charge to become transversely non-linear, leading to emittance growth \cite{LindstromThevenet_JINST_2022}.
In short, to preserve the beam quality of a trailing bunch \cite{Lindstrom_NatCommun_2024}, re-acceleration must be avoided, which ultimately limits driver-energy depletion.

In this Letter, we experimentally observe re-acceleration of energy-depleted driver electrons for the first time, using two separate diagnostics, and demonstrate a driver-to-plasma energy-transfer efficiency of (57~$\pm$~3)\% at the re-acceleration limit by accurately measuring and reconstructing the full energy spectrum.

The experiment was performed in the FLASHForward plasma accelerator at DESY \cite{DArcy_PTRSA_2019}, using 501~MeV electron bunches from the FLASH free-electron-laser linac \cite{FLASH}.
The electron bunches with charge ($636\pm8$)~pC were compressed using two magnetic chicanes to a root-mean-square (rms) bunch length of ($83\pm3$)~fs, as measured by an X-band transverse-deflection structure \cite{Craievich_PRAB_2020, Marchetti_SciRep_2021}. 
Nine quadrupole magnets were used to focus the beam at the entrance of the plasma accelerator to approximately  $17$~{\textmu}m rms transverse size; this was estimated based on upper and lower bounds on divergences, emittances and beta functions (the latter measured with the technique in Ref.~\cite{Lindstrom_PRAB_2020}), and from comparisons between experiments and simulations.
The plasma was created using a high-voltage discharge through a 195-mm-long sapphire capillary~\cite{Butler_PRL_2002} filled with argon gas.
Doping argon with 3\% hydrogen allows the measurement of the plasma-electron density via optical-emission spectroscopy using the radially averaged H-alpha line~\cite{Garland_RevSciInst_2021}. The density was observed to decay approximately exponentially with a half-life of ($2.6\pm0.1$)~{\textmu}s (between 5 and 12~{\textmu}s after the initial discharge).
The density experienced by the beam was adjusted by varying the discharge-trigger timing.
Moreover, the longitudinal plasma-density profile is expected to have Gaussian-like plasma-density ramps due to expulsion on this timescale, as measured in Ref.~\cite{Garland_RevSciInst_2021}.
To estimate the energy deposition by the drive bunch in the plasma, optical-emission light was recorded with a CMOS camera from one side of the plasma cell and integrated over 24~{\textmu}s starting a few {\textmu}s after the beam interaction.
Downstream of the plasma accelerator, an imaging energy spectrometer consisting of five quadrupoles and a vertically dispersive dipole magnet enabled the measurement of 98\% of the energy spectrum (from 12 to 501~MeV) on a gadolinium-oxysulfide screen recorded using three CMOS cameras with partly overlapping fields of view.

To measure the energy depletion of the driver and observe the onset of re-acceleration, the energy spectrum of the electrons must be measured while varying the energy loss. 
For a given driver, the depletion can be increased in two ways: by lengthening the plasma traversed by the bunch (as shown in Fig.~\ref{fig:1_simulation}); or by increasing the plasma density, which raises the amplitude of the decelerating field and thus the energy-transfer rate to the plasma.
Only the latter is realizable in the experimental setup described above. 
However, increasing the plasma density leads to a decrease in the plasma wavelength: for all electrons to lose energy, the driver must be confined to only the decelerating region of the wake (i.e., be shorter than half a plasma wavelength).
For the bunch used in this experiment, the plasma density was therefore restricted to a maximum of $\sim$$4\times10^{16}$~cm$^{-3}$.
No acceleration was observed beyond the initial driver energy.
Furthermore, the rms transverse size of the focused beam was smaller than the plasma-wake radius (approximately equal to the plasma skin-depth) for the entire range of densities.

\begin{figure}[t]
	\centering\includegraphics[width=1\linewidth]{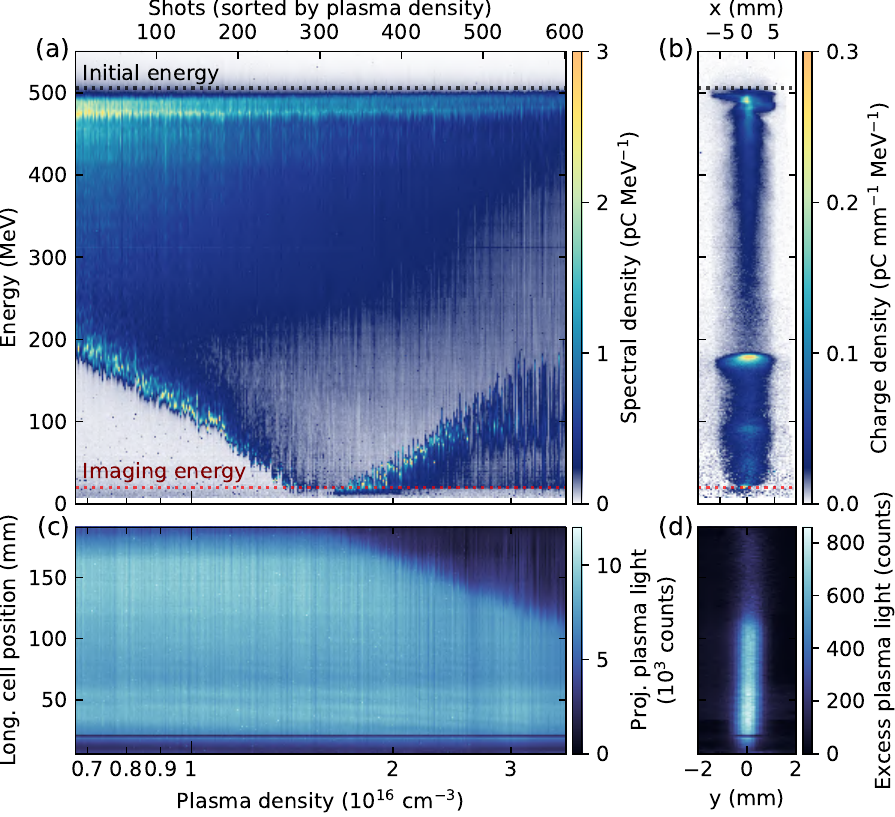}
	\caption{(a) Energy spectra shown for a scan of plasma density, where (b) shows an example spectrometer image (shot 594, corresponding to a density of $3.5 \times 10^{16}$~cm$^{-3}$). The spectrometer optics were set to point-to-point image 20~MeV electrons (red dotted line). (c) Plasma-emission light along the plasma cell from 6 to 191~mm downstream of the plasma entrance for the corresponding shots in (a), showing images projected onto the longitudinal axis (i.e., along the plasma cell); the image in (d) represents the same shot as (b).}
    \label{fig:2_reacceleration}
\end{figure}

Figure~\ref{fig:2_reacceleration}(a) shows the energy spectrum of the driver as a function of plasma density.
At the lowest density ($\sim$$7\times10^{15}$~cm$^{-3}$), driver electrons lose up to 300~MeV; dropping from 501~MeV to 200~MeV.
As the plasma density increases, the electron-energy loss increases until $\sim$$1.5\times 10^{16}$~cm$^{-3}$, where the lowest-energy electrons fall below the measurement limit of the spectrometer ($\sim$12 MeV), where particles become non-relativistic.
This density threshold agrees with estimates derived from well-known models~\cite{Lu_PhysPlasmas_2006}.
At higher densities, a distinct peak appears in the spectrum, which moves to higher energy as the plasma density increases.
This observation is consistent with re-acceleration of energy-depleted electrons, as seen in Fig.~\ref{fig:1_simulation}(k). 
At the maximum density explored here ($\sim$$4\times10^{16}$~cm$^{-3}$), some electrons are re-accelerated to nearly 200~MeV, 40\% of the initial energy.
A spectrometer image showing such electrons can be seen in Fig.~\ref{fig:2_reacceleration}(b).

Additional evidence of re-acceleration is provided by the plasma-emission light, whose intensity at a given longitudinal position correlates with the net energy locally deposited by the driver electrons \cite{Oz_AIPConfProc_2004,Boulton_PRL_2022}. 
Figure~\ref{fig:2_reacceleration}(c) shows the longitudinal variation of \textit{excess} plasma-emission light (i.e., subtracting the light without beam--plasma interaction) along the plasma cell for the shots corresponding to the scan in Fig.~\ref{fig:2_reacceleration}(a). 
The non-uniform longitudinal structure of the emission, which is largely independent of plasma density, probably originates from diagnostic artifacts (reflections, vignetting, and other attenuations).
To maintain a constant background throughout the plasma-density scan, the camera was triggered at a constant delay with respect to the discharge. 
This changes the delay between the beam arrival and the image acquisition. 
Thus, while more energy is deposited at higher densities, there is more time for it to dissipate before the camera acquisition begins. 
These opposing effects explain why the plasma density only weakly affects the light intensity. 

We observe that for plasma densities below $\sim$$1.5 \times 10^{16}$~cm$^{-3}$, the longitudinal plasma-light profile is approximately unchanged, whereas, for higher densities, a distinct drop in signal appears at the downstream end, which moves upstream as the density increases.
Such a drop in signal is consistent with re-acceleration for two reasons. 
First, less energy is deposited in the plasma wake due to decreased driver charge. 
Secondly, energy is extracted through beam loading by the re-accelerated charge~\cite{Boulton_PRL_2022}.
Moreover, the beam-plasma interaction length at which the first electrons become non-relativistic and slip in phase is localized by this drop, which can be used to estimate the average decelerating and accelerating gradients.
For example, Figs.~\ref{fig:2_reacceleration}(b) and \ref{fig:2_reacceleration}(d) show that electrons with an initial 501~MeV were energy depleted over the first 115~mm of interaction (4.3~GV/m), while the re-accelerated electrons gained 180~MeV in the remaining 80~mm of plasma (2.3 GV/m).

Determining the \textit{driver-depletion efficiency} (fraction of the driver energy deposited into the plasma wake) requires accurately measured energy spectra.
While the spectra in Fig.~\ref{fig:2_reacceleration}(a) are sufficient to draw the qualitative conclusion that re-acceleration has been observed, they are not quantitatively correct across the full energy range for two reasons: improper point-to-point imaging away from the imaging energy [20~MeV in Fig.~\ref{fig:2_reacceleration}(a)]; and some charge loss in transport from the plasma cell to the spectrometer---a problem aggravated at lower energies.
If the energy of the lost charge is unknown, it will contribute to the uncertainty of the measured driver-depletion efficiency.

\begin{figure}[t]
	\centering\includegraphics[width=1\linewidth]{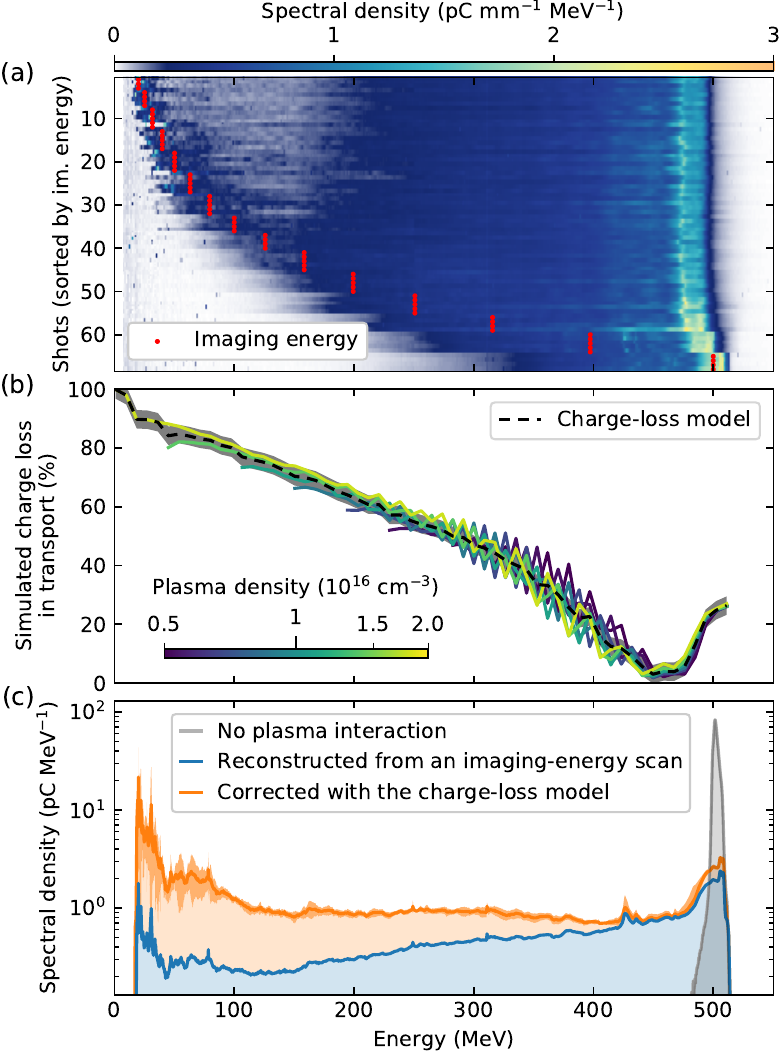}
	\caption{(a) Driver spectra measured at a plasma flattop density of $1.4\times10^{16}$~cm$^{-3}$ for a scan of imaging energies (red points).
    (b) The fractional charge loss versus energy for PIC-simulated bunches, as caused by transport losses between the plasma-cell exit and the spectrometer screen, is approximately independent of plasma density.
    The average serves as a charge-loss model (dashed line) with an average rms error of 3.1\% (gray band).
    (c) The reconstructed average spectrum (blue area) shows significant energy loss compared to no plasma interaction (gray area), but only accounts for 39\% of the initial charge. The spectrum corrected by the charge-loss model (orange area), with a band indicating the rms error, accounts for all the charge.}
    \label{fig:3_reconstruction}
\end{figure}

Point-to-point imaging is used to cancel the spectral distortion caused by angular offsets from plasma-induced kicks or divergence, but its useful energy range is limited by the chromaticity of the imaging quadrupoles.
To overcome this, an average spectrum can be measured via a scan of imaging energies [Fig.~\ref{fig:3_reconstruction}(a)] and subsequently combining the appropriately imaged regions of each spectrum [Fig.~\ref{fig:3_reconstruction}(c)].
This experiment achieved  a beam--plasma interaction sufficiently stable to carry out such a multi-shot measurement, thereby
allowing the spectral density of the beam that reaches the spectrometer screen to be approximated.

The spectrum of the charge lost before reaching the spectrometer screen is not measurable but can be estimated with a semi-empirical model.
Simulations (see Fig.~\ref{fig:1_simulation}), which approximate the actual experiment, indicate that charge is not lost during the plasma interaction, but rather in the post-plasma transport due to a narrow (5~mm radius) beam-pipe aperture with a maximum transmissible divergence during point-to-point imaging of $\pm7.5$ and $\pm2.0$~mrad in the horizontal and vertical planes, respectively.
As driver electrons lose energy, their divergence is expected to grow for three main reasons: since the longitudinal momentum ($p_s$) is reduced, the divergent angle ($x' = p_x/p_s$) for a given transverse momentum ($p_x$) will increase \cite{Floettmann_2003}; 
as the energy drops, the relative focusing strength increases (i.e., the matched beta function becomes smaller), causing a larger divergence \cite{Ariniello_PRAB_2019}; 
lastly, that fraction of the beam ahead of the wake blowout regime experiences non-linear focusing fields leading to emittance growth \cite{LindstromThevenet_JINST_2022}, producing higher divergence. 
While the first of these effects can be estimated analytically, the others depend on parameters non-trivial to measure to the required accuracy. 
Instead, these effects were estimated using PIC simulations based on measured input parameters, by propagating the simulated beam particles exiting the plasma to the spectrometer using energy-dependent transport matrices; particles that diverged beyond the radius of the beam-pipe were removed.
Fig.~\ref{fig:3_reconstruction}(b) shows the fractional charge lost as a function of energy for different plasma densities without electron re-acceleration. 
Here, the fractional charge loss is seen to be mainly a function of particle energy and only weakly dependent on plasma density.
As such, by averaging over density (to reduce the effect of betatron oscillations), a \textit{charge-loss model} that predicts the fraction of charge that reaches the spectrometer at a given energy has been constructed (see Supplementary Material).
Dividing the measured reconstructed spectrum by this fraction allows estimating the spectrum at the plasma exit, as shown in Fig.~\ref{fig:3_reconstruction}(c).

\begin{figure}[!t]
	\centering\includegraphics[width=1\linewidth]{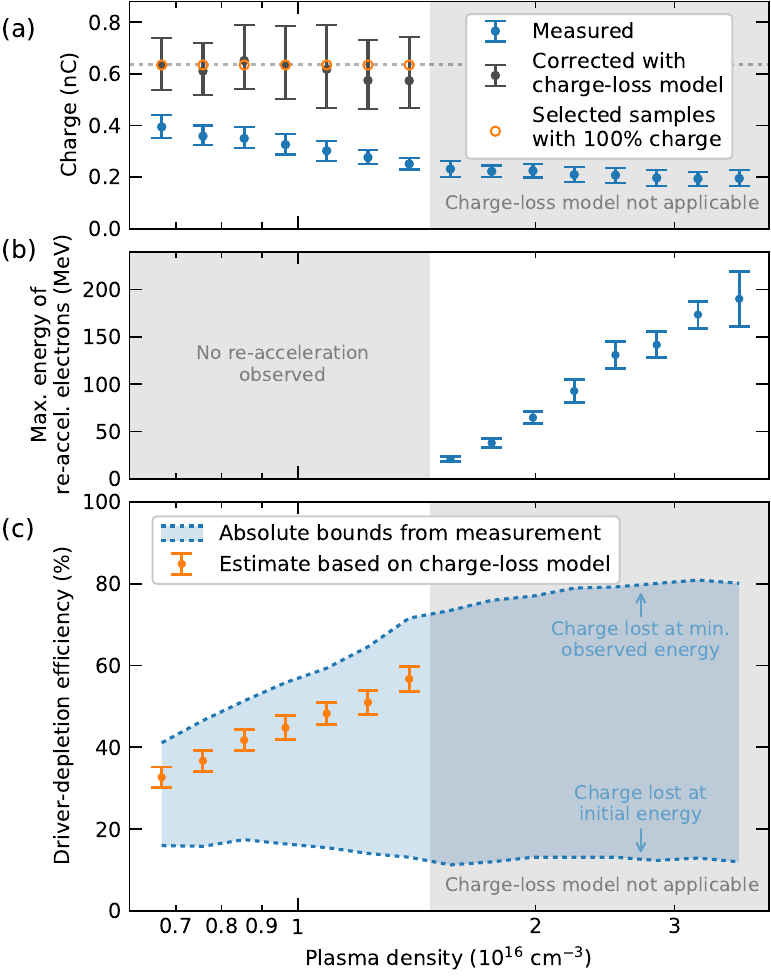}
    \caption{(a) For a scan of plasma density, the beam charge was measured upstream of the plasma cell using a beam-position monitor (dashed line) and downstream using the spectrometer screen, with and without correction (blue and gray error bars; representing the median and central 68-percentile range from sampling the charge-loss model and spectra within their errors).
    For the depletion estimate, we select the samples at each density that reconstruct all the lost charge (orange circles).
    (b) The maximum energy of the re-accelerated charge is shown (blue error bars; mean and rms) unless no such charge was observed.
    (c) The resulting driver-depletion efficiency is calculated based on the charge-corrected energy spectra (orange error bars; mean and rms of the samples reconstructing all the lost charge). Lower and upper bounds from the measurement are also shown (blue dashed lines). Plasma densities have an uncertainty of 7\% rms.
    }
    \label{fig:4_depletion}
\end{figure}

Figure~\ref{fig:4_depletion} shows a scan of plasma density over the same range as in Fig.~\ref{fig:2_reacceleration}, for which the spectrum was measured with an imaging-energy scan at each density step (as in Fig.~\ref{fig:3_reconstruction}).
Using the measured spectra, we determine the lower and upper bound of the driver-depletion efficiency by assuming that the missing charge experienced no energy loss, or the maximum observed energy loss, respectively.
To reduce the uncertainty, we apply the model described above to each scan step, with Fig.~\ref{fig:4_depletion}(a) showing the charge of the measured and corrected spectra along the scan. 
The model is only applicable to spectra without re-acceleration [densities below $1.5\times10^{16}$~cm$^{-3}$; Fig.~\ref{fig:4_depletion}(b)], as re-accelerated electrons have a different and unaccounted for divergence distribution. 
The driver-depletion efficiency shown in Fig.~\ref{fig:4_depletion}(c) can be seen to be compatible with a power-law increase with plasma density.
The value of the exponent depends on details of the plasma and beam parameters, but would be expected to be $\ge 0.5$~\cite{Lu_PhysPlasmas_2006}.
For the highest density without re-acceleration, the corrected spectrum reduces the uncertainty of the driver-depletion efficiency from $\pm29$\% to $\pm3$\%; the resulting estimate of the maximum efficiency is (57~$\pm$~3)\%.

Combining experimentally the maximum driver-depletion (driver-to-plasma) efficiency demonstrated here with the driver-production (wall-plug-to-driver) efficiency envisaged for CLIC (55\%~\cite{CLIC_2012}) and the previously demonstrated energy-extraction (plasma-to-beam) efficiency (42\%~\cite{Lindstroem_PRL_2021}), the overall wall-plug-to-beam efficiency of such a beam-driven plasma accelerator would be 14\%---similar to state-of-the-art radio-frequency accelerators (not including energy-recovery linacs \cite{Schliessmann_NatPhys_2023}).
In principle, the driver-depletion efficiency can be further increased by precise shaping of the driver-current profile, leading to a more uniform decelerating field across the drive bunch \cite{Lotov_PhysPlasmas_2005, Loisch_PRL_2018, Roussel_PRL_2020}, allowing more energy to be transferred to the trailing bunch before re-acceleration begins. 
However, simultaneously achieving high driver-to-plasma and plasma-to-beam efficiency can be complicated by transverse instabilities, such as hosing~\cite{Whittum_PRL_1991} or beam-breakup~\cite{Panofsky_RevSciInst_1968}, which are expected to become more severe at longer propagation distances and higher plasma-to-beam efficiencies~\cite{Lebedev_PRAB_2017}.
Promising concepts for suppressing these instabilities have been proposed~\cite{Mehrling_PRL_2017,Benedetti_PRAB_2017,Lehe_PRL_2017,Mehrling_PRL_2018,Ossa_PRL_2018,Mehrling_PRAB_2019, Diederichs_PhysPlasmas_2022}.
The result presented in this Letter is a major step toward high-overall-efficiency plasma accelerators, promising more environmentally responsible particle accelerators.

\begin{acknowledgments}
    The authors would like to thank M.~Dinter, S.~Karstensen, S.~Kottler, K.~Ludwig, F.~Marutzky, A.~Rahali, V.~Rybnikov, A.~Schleiermacher, the FLASH management, and the DESY FH and M divisions for their scientific, engineering and technical support. This work was supported by Helmholtz ARD and the Helmholtz IuVF ZT-0009 programme, the Research Council of Norway (NFR Grant No.~313770), as well as the Maxwell computational resources at DESY. The authors gratefully acknowledge the Gauss Centre for Supercomputing e.V. (www.gauss-centre.eu) for funding this project by providing computing time through the John von Neumann Institute for Computing (NIC) on the GCS Supercomputer JUWELS at Jülich Supercomputing Centre (JSC).
\end{acknowledgments}


\end{document}


\maketitle


\section{Validity of the quasi-static approximation}

\begin{figure}[!b]
    \centering\includegraphics[width=\linewidth]{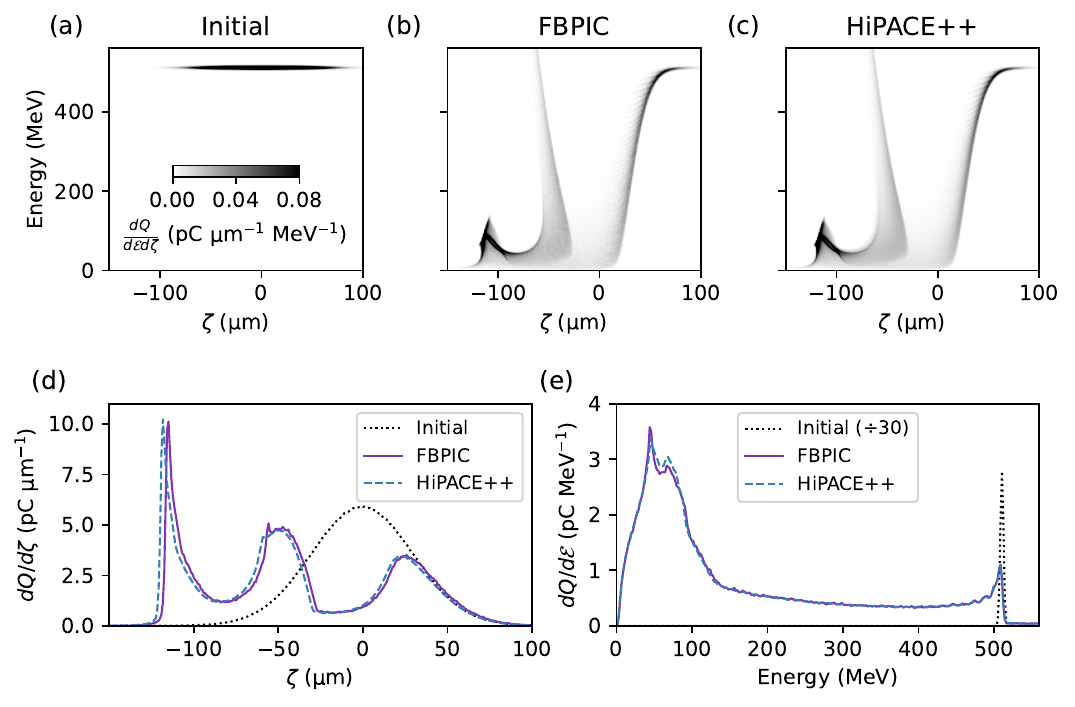}
    \caption{Longitudinal phase-space of the initial beam (a) and final beam from simulations using HiPACE++ (b) and FPBIC (c) along the co-moving variable $\zeta = s - ct$, with $s$ being the longitudinal coordinate, $t$ the time, and $c$ the speed of light in vacuum.
    The flattop plasma density in the simulations is $3.7\times 10^{16}$~cm$^{-3}$.
    The longitudinal charge distribution (d) and the energy spectrum (e) show excellent agreement between the different PIC codes.
    }
    \label{fig:supp:simulation_check}
\end{figure}

The particle-in-cell (PIC) code HiPACE++~\cite{Diederichs_2022} relies on the quasi-static approximation, which might be inaccurate in the presence of non-relativistic and re-accelerated electrons. 
To legitimate the usage of a quasi-static code, we performed a benchmark of the simulation used in Fig.~1 with respect to the full electromagnetic PIC code FBPIC~\cite{Lehe_CPC_2016}. 
As can be seen in the supplementary Fig.~\ref{fig:supp:simulation_check}, re-acceleration is fully captured by HiPACE++ for this range of parameters.

\section{Charge-loss model and estimation of the error in depletion efficiency}

Simulations recreating the experimental setup suggest that the charge is lost in transport from the plasma exit to the spectrometer diagnostic.
As the charge loss is mainly dependent on the energy of the electrons [see Fig.~3(b)], and largely independent of the plasma density [within the scan range; see Fig.~3(b)], we use an average from the different simulated densities, which averages out varying charge loss from betatron oscillations.\\

As it is uncertain how the error in the model is distributed throughout the spectrum, we assume a constant error throughout the energy range: the average of the root-mean-square (rms) at each energy slice, which is 3.1\%. 
To assess how this affects the charge of the corrected spectra, we use Monte Carlo (MC) sampling. 
We sampled different charge-loss models ($\pm$3.1\%, normally distributed) and, simultaneously, different spectra (normally distributed within the measured rms variation), and show the charges of the corrected spectra in Fig.~4(a): the error bars (dark gray) represent the median and central 68-percentile range.\\

Next, since we know the charge of the incoming beam (from an upstream measurement) we can select, for each plasma density, all the samples that reconstruct all the missing charge [depicted with the orange circles in Fig.~4(a)]. 
An example corrected spectrum is shown in orange in Fig.~3(c), where the bands show the rms variation around the average reconstructed spectrum.
With a corrected spectrum that accounts for all the incoming charge, we can estimate the driver-depletion efficiency [orange error bars in Fig.~4(c)].\\

The uncertainty of this efficiency estimate can be found from the MC samples.
Since the model and the measured spectra are each sampled independently within their errors (normally distributed), different combinations of model and spectrum samples can result in 100\% reconstruction. Using this subset of samples, the corresponding variation in depletion efficiency (rms) can be used as the depletion-efficiency uncertainty. \\

To account for possible correlations within the spectrum and model errors, we repeat the process described above for energy bins of different widths ranging from 1~MeV to 500~MeV. 
The error bars in Fig.~4(a) and (c) represent the largest of these errors, which is found to be at approximately 275~MeV bin width. 
This constitutes an upper bound to this error. \\